\documentclass[aps, prl, floatfix, superscriptaddress, showpacs]{revtex4-1}

\usepackage{graphicx}
\usepackage{color}
\usepackage{calc}
\usepackage{array}
\usepackage{graphicx}
\usepackage{amsmath, amssymb}
\usepackage{hyperref}

\begin{document}

\title{Prediction of superconducting transition temperatures of heterostructures based on first principles}

\author{G{\'{a}}bor \surname{Csire}}%
\affiliation{Institute for Solid State Physics and Optics,
             Wigner Research Centre for Physics, Hungarian Academy of Sciences, \\
             PO Box 49, H-1525 Budapest, Hungary}
\affiliation{Department of Physics of Complex Systems, E{\"{o}}tv{\"{o}}s University,
             H-1117 Budapest, P{\'a}zm{\'a}ny P{\'e}ter s{\'e}t{\'a}ny 1/A, Hungary}

\author{J{\'o}zsef Cserti}
\affiliation{Department of Physics of Complex Systems, E{\"{o}}tv{\"{o}}s University,
             H-1117 Budapest, P{\'a}zm{\'a}ny P{\'e}ter s{\'e}t{\'a}ny 1/A, Hungary}

\author{Istv\'{a}n T\"{u}tt\H{o}}
\affiliation{Institute for Solid State Physics and Optics,
             Wigner Research Centre for Physics, Hungarian Academy of Sciences, \\
             PO Box 49, H-1525 Budapest, Hungary}

\author{Bal{\'a}zs {\'U}jfalussy}
\affiliation{Institute for Solid State Physics and Optics,
             Wigner Research Centre for Physics, Hungarian Academy of Sciences, \\
             PO Box 49, H-1525 Budapest, Hungary}



\begin{abstract}

In this paper we present material specific calculations
of superconductor -- normal metal heterostructures using density functional theory.
In particular, we calculate the quasiparticle spectrum of different normal metal overlayers on a Nb(100) host.
We find that the Andreev reflection leads to the formation of momentum dependent quasiparticle bands in the normal metal.
As a consequence, the spectrum has a strongly momentum dependent induced gap.
From the thickness dependence of the gap size we calculate the superconducting critical temperature
of Au/Nb(100) heterostructures where we find very good agreement with experiments. Moreover, predictions are
made for similar heterostructures of other compounds.
\end{abstract}

\maketitle


The experiments of Yamazaki et al.~\cite{Yamazaki1, Yamazaki2} poses a great challenge to the theory of superconductivity.
In this experiment they studied the critical temperature  of 
a very thin layer of gold grown on top of a thicker Nb host.
While the thickness of the Nb sample (288~{\AA}) was in the order of the superconducting coherence length (380~{\AA}),
the gold overlayer was only a couple of atomic layers thick.
It was found that the superconducting transition temperature is significantly lowered in this composite
compared to the bulk, and it decays monotonically as the thickness of the gold overlayer increases.
While this appears to be a plausible consequence at first sight, it is surprising that the entire composite is superconducting.
The explanation of these experimental findigs based on  the standard theory of superconductivity is very difficult, because within the BCS 
(Bardeen-Cooper-Schriffer) theory~\cite{BCS} it is not easy to describe inhomogeneous systems.
However such systems can be treated efficiently in terms of the Bogoliubov-de Gennes (BdG) equations~\cite{Bogo,deGennes}, 
especially when they are reformulated within the framework of density functional theory~\cite{OGK,Suvasini,Luders1,Luders2,Csire}
leading to the so called Kohn-Sham-Bogoliubov-de Gennes (KS-BdG) equations.
In principle these equations are able to describe systems with inhomogeneities in the pairing potential.

In inhomogeneous systems, such as  multilayers, it is expected that the Andreev reflection~\cite{Andreev64,BTK} 
leads to the formation of certain bound states, often referred to as Andreev bound states.
Our main aim in this paper is to calculate the dispersion relation of Andreev states.
We also verify that this effect enables the transport of supercurrents through non-superconducting materials, and
is the key to predict the superconducting transition temperature in the whole system.

In what follows we numerically solve the KS-BdG equations, as it was described in Ref.~\cite{Csire}, for an 
Au/Nb(100) overlayer system.  For varying Au thicknesses we calculate the quasiparticle electronic structure and obtain the superconducting gap for each thickness.
Finally a method is developed to calculate the critical temperature from the gap and applied to several overlayer systems with different material constituents.

The first step down this road is to construct a realistic model of the interface lattice structure.
Such a model can be built up from two-dimensional translational invariant atomic layers as follows.
We divide the system into three regions: (i) a semi-infinite bulk (Nb); (ii) the interface region
that -- in our case -- consists of six superconducting layers (Nb), various number of normal metal layers and
three layers of empty spheres; (iii) and a semi-infinite vacuum. This division of space is typical in Screened KKR \cite{Szunyogh}
calculations in overlayer systems.
Although the thickness of the Nb layers in the experiment is not strictly semi-infinite, it is thick enough to be 
approximated by a semi-infinite bulk system. We use six Nb layers to mimic the transition between the semi-infinite bulk and the 
overlayer. The Nb has the body-centered cubic (bcc) crystal structure with a lattice parameter of $a = 3.3$~\AA.
We assumed the realistic face-centered cubic (fcc) crystal structure for the gold overlayers, and the termination of the material layers were
modeled by a semi-infinite vacuum. We also neglected imperfections at the interface, like interlayer relaxations and intermixing.

In such systems, where translational invariance is preserved only parallel to the interface, the quasiparticle spectrum  is obtained
as a function of a two dimensional momentum vector $\vec{k}_{||}$ . To visualize this spectrum, it is customary 
to calculate the Bloch spectral function (BSF).
In two dimensionally translational invariant systems 
the BSF for layer $I$ is defined as
$A_B^I(\varepsilon, \vec{k}_{||})= -\frac{1}{\pi} \mathrm{Im} G_{II}(\varepsilon,\vec{k}_{||})$,
where $G_{II}$ is the layer projected Green function (see Ref.~\cite{Csire}).
Since the BSF is equivalent to the quasiparticle spectrum, drawing a contour plot of
the BSF as a function of energy along specified directions of $\vec{k}_{||}$ is a powerful tool to visualize the quasiparticle states.
Evidently, in a layered system this contour plot can be done for each layer.

\begin{figure}[hbt!]
   \includegraphics[width=0.9\linewidth]{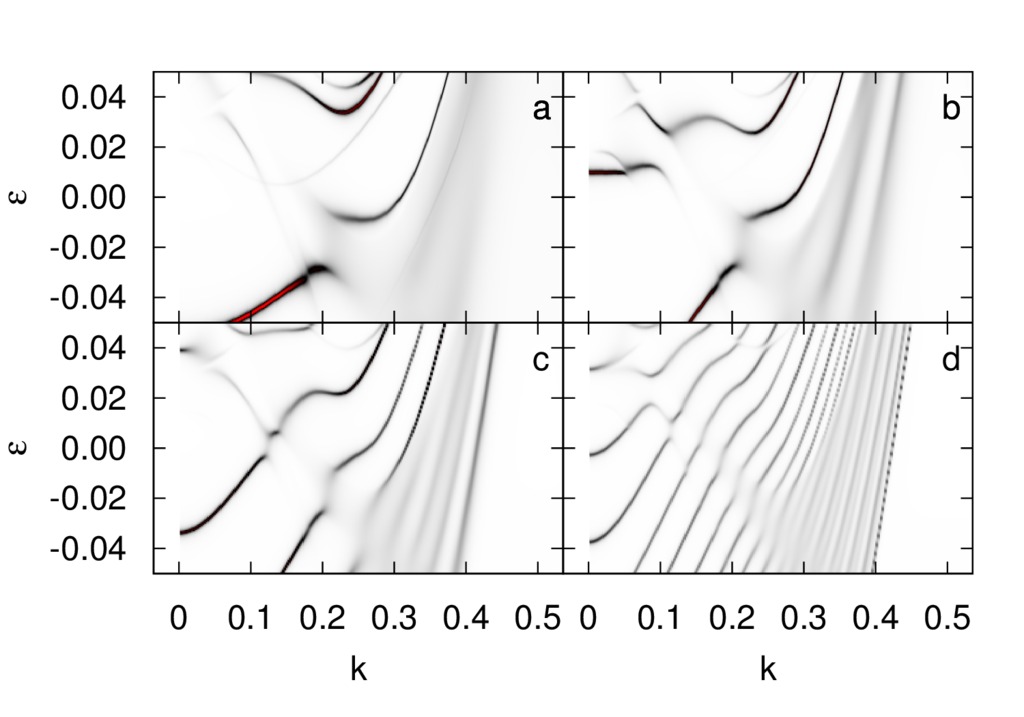}
   \caption{\label{fig:normal1}%
            Contour plot of the spectral function of Au/Nb(001) (normal state band structure)
            from the "middle" of the gold layers for different thicknesses of the gold:
            3 Au layers (a) and 6 Au layers (b)
            9 Au layers (c) and 24 Au layers (d). }
\end{figure}

In order to understand the quasiparticle spectrum in the superconducting state better, first we performed first principles calculations 
just in the normal state, by simply solving the Screend KKR equations~\cite{Szunyogh}  for the
Au/Nb(001) overlayer system. In Fig.~\ref{fig:normal1} we show the contour plot of the BSF
for a layer that we considered to be in the "middle" of the samples of various Au layer thicknesses.
The plots are restricted in energy to the range of [-0.05~Ry, 0.05~Ry] (later we will choose
${\Delta}_{Nb}$, the superconducting gap to equal this value, and solve the BdG equations within this energy range).
It can be seen from  Fig.~\ref{fig:normal1} 
that for energies where the DOS in the bulk Nb is low, the states in the Au are confined, as they can not scatter into the Nb,
and on the other side the system is limited by vacuum. In regions where the DOS is high in the Nb, the states in the Au
are smeared out, as here the electrons can scatter more easily into the other side of the interface. The confined states
in the Au can be regarded as quantum-well (QW) states. 

Now let us consider the Nb in the superconducting state and solve the first-principles BdG equations as
described in Ref.~\cite{Csire}. In this formalism each layer is characterized by an additional, potential-like parameter,
the  superconducting gap ${\Delta}$. To solve
the BdG equation we assumed that all Nb layers show the same superconducting gap ${\Delta}_{Nb}=0.05$~Ry. This is a much larger value
than the gap observed experimentally~\cite{Pronin}, in order to make it easier to visualize the quasiparticle spectrum. Later, when we will calculate the 
superconducting transition temperature, we will repeat the calculation for a more realistic $\Delta$. We also assumed, that the
superconducting gap is zero in the Au layers, namely we set ${\Delta}_{Au}=0.0$~Ry, as Au is not superconducting at all in the bulk.

\begin{figure}[hbt!]
   \includegraphics[width=1\linewidth]{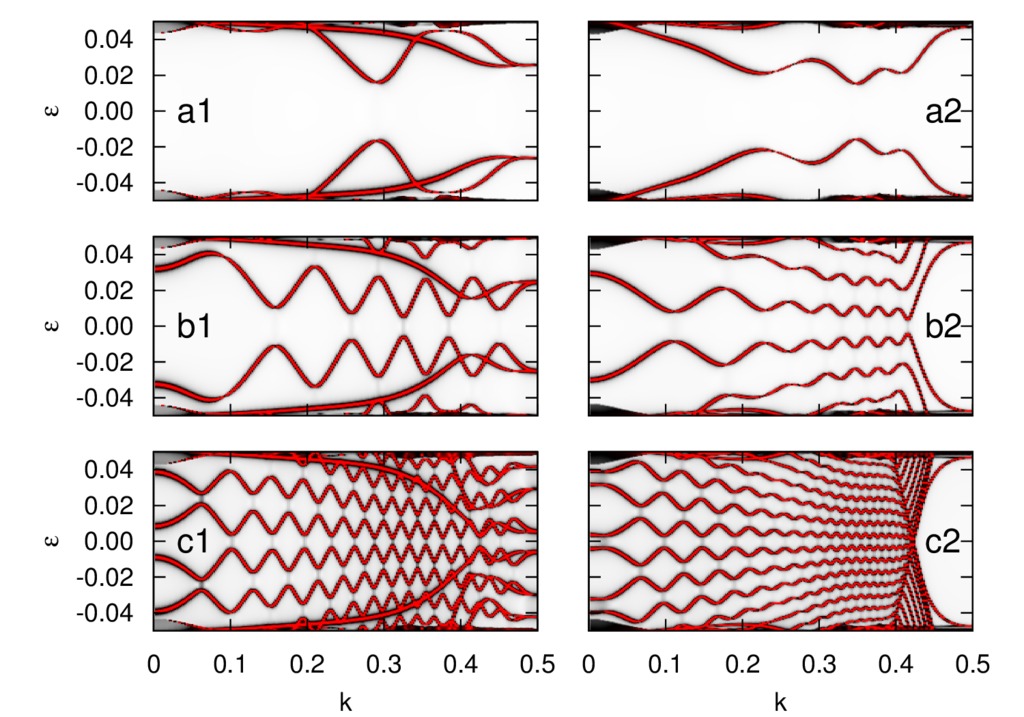}
   \caption{\label{fig:supra_band}%
            Contour plot of the BSF summed up for all Au layers of Au/Nb(001) in the case of different different thicknesses of the Au:
            3 bcc Au layers (a1), 9 bcc Au layers (b1),
            24 bcc Au layers (c1), and
            3 fcc Au layers (a2), 9 fcc Au layers (b2),
            24 fcc Au layers (c2). The momentum vector k is shown along the $k_x=k_y$ line in the 2D Brillouin zone.}
\end{figure}

Based on the symmetry of the BdG equations~\cite{deGennes} it is known, that in bulk superconductors the quasiparticle states can be obtained
from the electronic states by mirroring the band structure of the normal state to the Fermi energy and opening up a gap.
In the case of the Au/Nb(100)  overlayer system the situation is entirely different, because the system is described by an inhomogeneous $\Delta$ which, additionally, 
equals to zero in some layers. Therefore the main question is how  the niobium host affects 
the quasiparticle spectrum of the gold overlayers.
In Fig.~\ref{fig:supra_band} we plot the BSF summed up for all Au layers in the case of different total number of gold layers
(right hand side of the figure).
To illustrate the role of the crystal structure, on the left hand side of the figure we also show results of the same calculations 
performed by assuming bcc Au layers throughout the overlayer. This is not entirely unrealistic for very thin layers but most 
likely artificial for thicker ones, where the Au layers
most certainly grow in an fcc structure. Unfortunately there is very little known of the crystal structures of thin Au layers on Nb.

By comparing Fig.~\ref{fig:normal1}/a,c,d and Fig.~\ref{fig:supra_band}/a2,b2,c2, we can immediately conclude 
that the proximity of a superconductor in the studied heterostructures induces the expected mirroring of the electronic bands 
within the energy range of the Nb gap, and additionally opens up a
smaller gap at each band crossing. Although it is quite trivial, it is much easier 
to see the effect first in the case of the bcc Au,  fully described in \cite{Csire}.
One can also see from the figure that the size of the induced gaps in the gold is strongly $\vec{k}_{||}$-dependent,
however  - around the Fermi energy -  it is the same for all layers, and it decreases as a function of the overall thickness of the gold overlayers.
It should be noted as well that those regions of the spectrum which were more or less smeared out in the normal state
(visualized in the figure by weaker lines) now sharpened up.
This is the consequence of the opening of the superconducting gap in the Nb: the states where scattering
into - on the other side of the interface - was allowed before, now disappeared because of the gap.
We can also conclude from the figure that the quantum-well states, which we found to exist in the normal state band structure calculations,
become bound Andreev states originating from the Andreev scattering at the interface.
Finally, since Fig.~\ref{fig:supra_band} was obtained by summing up $A_B^I(\varepsilon, \vec{k}_{||})$, and this sum does not show any significant
broadening, it is clear that the quasiparticle spectrum is virtually identical in every interfacial layer.
In Fig.~\ref{fig:supra_band} one can also observe changes in the quasiparticle spectrum owing to different lattice structures.
In particular, more 'oscillations' can be seen in fcc gold overlayers compared to the bcc ones,
which is the consequence of more bands of the fcc gold in the normal state.

\begin{figure}[hbt!]
   \includegraphics[width=1\linewidth]{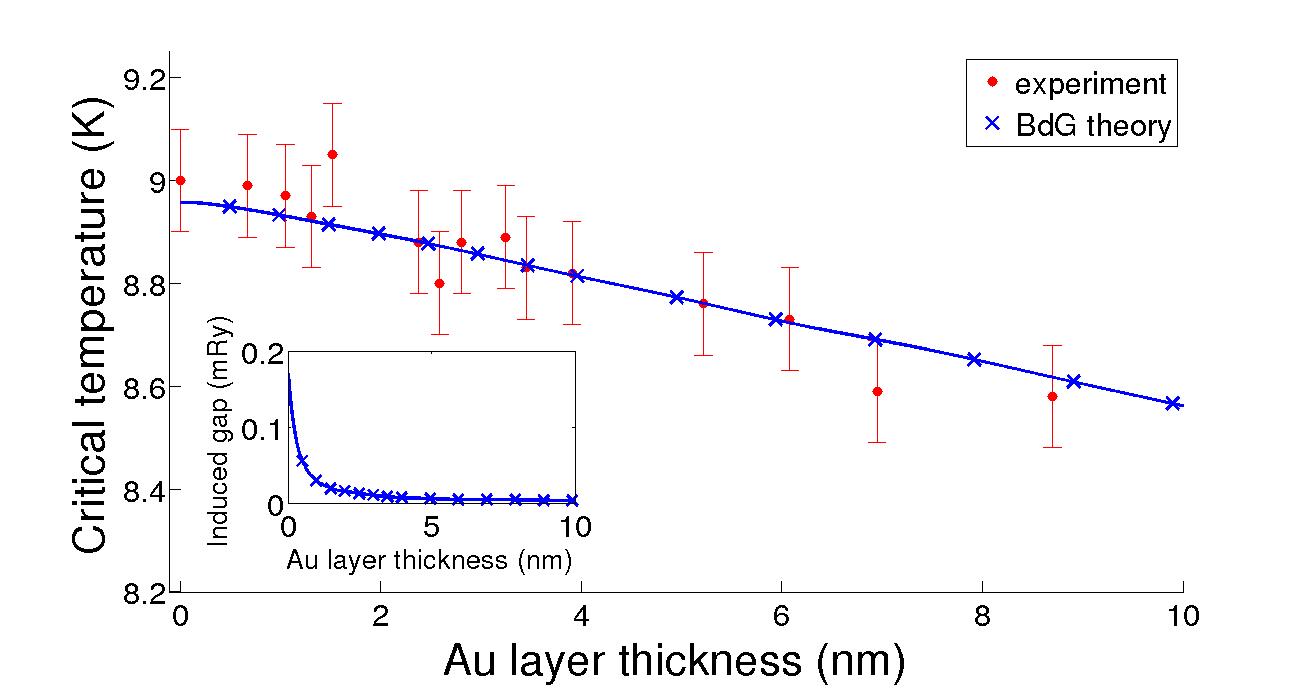}
   \caption{\label{fig:tc}%
            (Color online) The critical temperature as function of the thickness of the gold obtained from the McMillan formula.
            The points with the error bars are the experimental results from Ref.~\cite{Yamazaki2}.
            The inset plot shows the induced gap on the gold layers as function of the thickness of the gold.}
\end{figure}

We are now ready to calculate the superconducting transition temperature $T_c$.
In the strong-coupling limit (e.g. in the case of Nb) the $T_c$ is given by the McMillan formula~\cite{McMillan,Allen}
\begin{equation}
  T_c= \frac{\Theta_D}{1.45} \exp{\left(-\frac{1.04(1+\lambda_{eff})}{\lambda_{eff}-\mu^*(1+0.62\lambda_{eff})}\right) },
  \label{eq:McMillan}
\end{equation}
where $\Theta_D$ is the Debye temperature, $\mu^*$ is the dimensionless Coulomb pseudopotential
which describes the effect of the Coulomb repulsion.
For this two parameter we may assume the bulk values, justified by the fact that the niobium host is rather thick.
The parameter $\lambda_{eff}$ is an effective electron-phonon coupling parameter for the whole system.
Following the argument of de Gennes~\cite{Yamazaki2,dGformula} the simplest way to define such an effective electron-phonon interaction parameter
in an inhomogeneous system is by an averaging process described in the following way.
Let us interpret $D_{Au}(\varepsilon_F)t_{Au} /(D_{Nb}(\varepsilon_F)t_{Nb}
+ D_{Au}(\varepsilon_F)t_{Au})$ as a probability of finding an electron in the normal metal layers (Au) and
$D_{Nb}(\varepsilon_F)t_{Nb} / (D_{Nb}(\varepsilon_F)t_{Nb} + D_{Au}(\varepsilon_F)t_{Au})$ in the superconducting Nb layers.
Then the $\lambda_{eff}$ effective electron-phonon coupling can be defined as
\begin{equation}
\lambda_{eff}= \frac{ \lambda_{Nb} D_{Nb}(\varepsilon_F)t_{Nb}+ \lambda_{Au} D_{Au}(\varepsilon_F)t_{Au}}{D_{Nb}(\varepsilon_F)t_{Nb} + D_{Au}(\varepsilon_F)t_{Au}},
\end{equation}
where $t_{Nb}$ and $t_{Au}$ are the thickness of the Nb and Au layers, respectively,  and $D(\varepsilon_F)$ is the density of states
at the Fermi level.
While the density of states for the Nb can be approximated by the bulk value, the density of states in the gold should be defined as a layer average,
$D_{Au}(\varepsilon_F)=1/N\sum_{I=1}^N D_{Au}^I(\varepsilon_F)$, where $N$ is the number of the gold overlayers.
Since the whole niobium -- gold system is superconducting (a common $T_c$ obtained experimentally in Ref. \cite{Yamazaki1, Yamazaki2}),
Cooper pairs must exist throughout the whole system. Therefore, the existence of the superconducting gap in the Au layers can be viewed
in a way that the proximity of the superconductor induces an effective $\lambda_{Au}$ electron-phonon coupling in the gold,
which in turn can be obtained from the induced gap $\Delta'_{Au}$ according to the BCS gap equation:
\begin{equation}
  \lambda_{Au}=-\frac{1}{\log \left (\frac{\Delta'_{Au}}{2\hbar \omega_D^{Nb}} \right)} .
\end{equation}
This induced gap $\Delta'_{Au}$ can be read off from the quasiparticle spectrum, 
for which the whole calculation of the quasiparticle spectrum needs to be repeated with a more realistic value for $\Delta_{Nb}$,
as we mentioned earlier.  (For better visualization we used an artificially large $\Delta_{Nb}$ in Fig.~\ref{fig:supra_band}.)
While the BCS theory is valid only for the weak-coupling limit, nevertheless we still argue that the BCS gap equation may be applied here to the gold layers,
since the induced gap is much smaller than the gap in the Nb.
To obtain the induced gap, we always take the maximum of the induced gap around the Fermi energy (as a function of $\vec{k}_{||}$)  in the 2D Brillouin zone 
because it is the most characteristic of the strength of the electron-phonon coupling.
In summary, the $\Delta'_{Au}$ induced gap can be read off from the calculated quasiparticle spectrum
as the function of the thickness of the gold overlayers (see Fig. \ref{fig:tc} inset), and the critical temperature can be calculated
from the  McMillan-formula given by Eq.~(\ref{eq:McMillan}).
Figure \ref{fig:tc} shows the critical temperature as a function of the thickness of the Au overlayers obtained from the procedure described above 
together with the experimental results taken from the papers~\cite{Yamazaki1, Yamazaki2}.
One can clearly see that our theoretical results are in very good agreement with the experimental findings, all points are within the experimental errors.
Moreover, the slow decrease of the critical temperature as function of the Au overlayer thickness is well reproduced.
We should emphasize that our theory contains only one parameter, namely the critical temperature for bulk niobium.
While in principle this quantity can be calculated~\cite{Klein}, here for simplicity we take its experimental value.

\begin{figure}[hbt!]
   \includegraphics[width=0.9\linewidth]{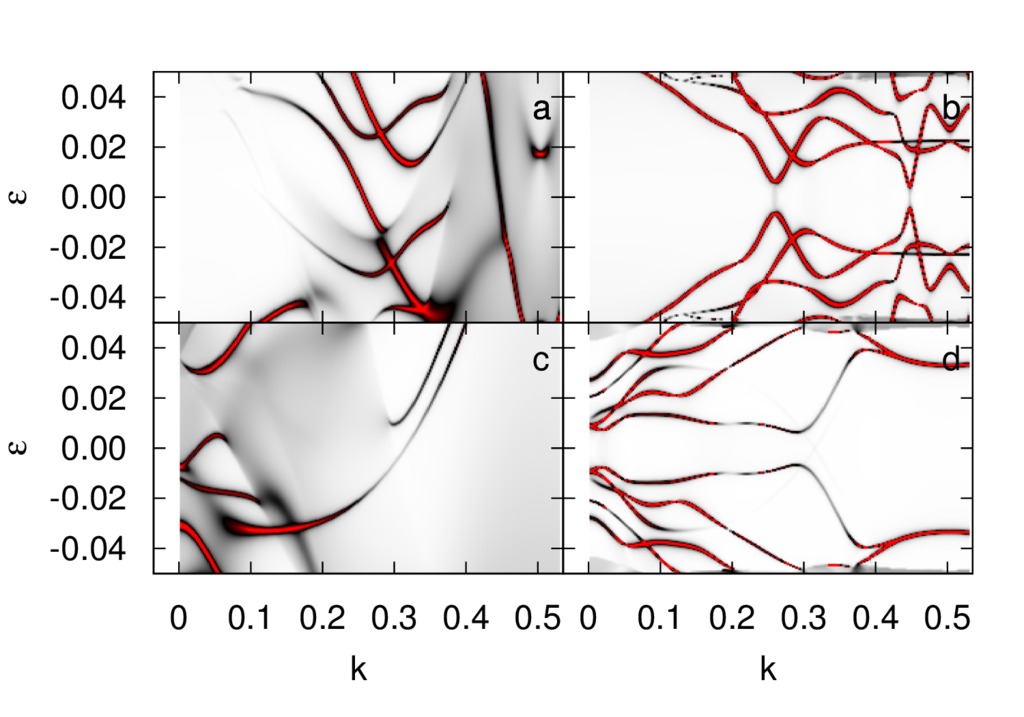}
   \caption{\label{fig:not_qw}%
            Contour plot of the spectral function from the "middle" of the 3 overlayers:
            (a) Al/Nb(001) (normal state band structure) and
            (b) Al/Nb(001) (quasiparticle spectrum);
            (c) Mo/Nb(001) (normal state band structure) and
            (d) Mo/Nb(001) (quasiparticle spectrum). }
\end{figure}

\begin{figure}[hbt!]
   \includegraphics[width=0.85\linewidth]{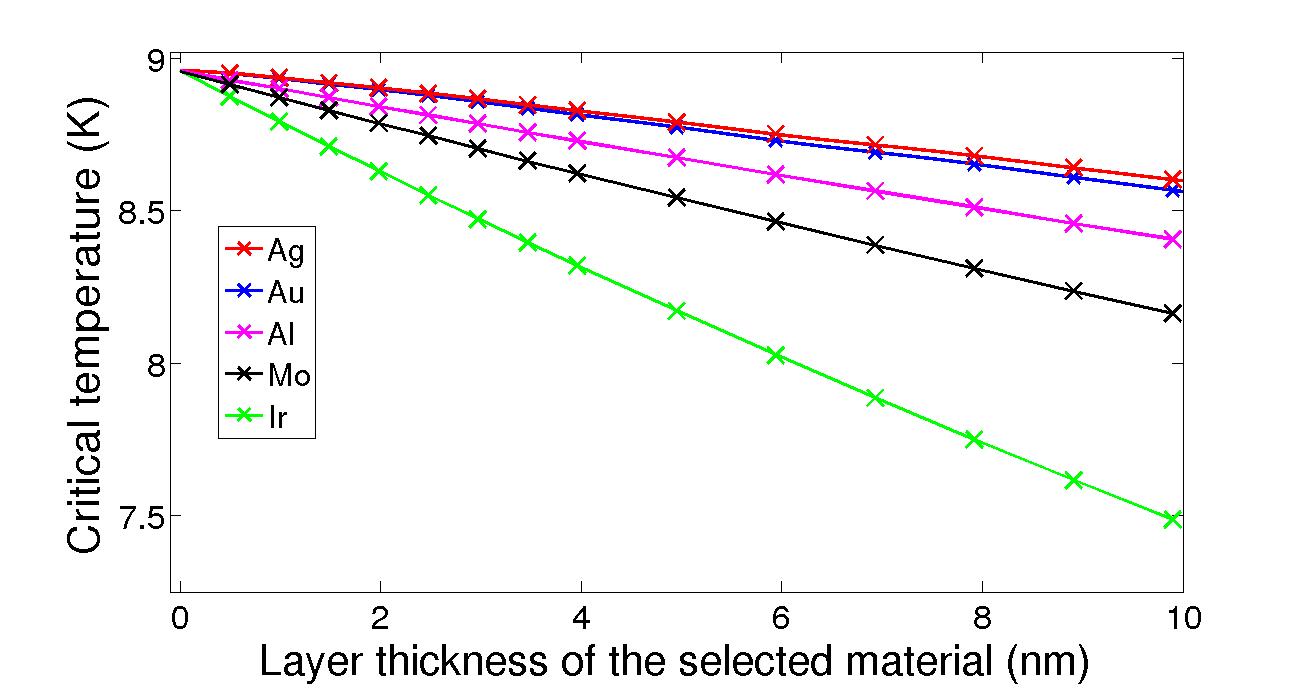}
   \caption{\label{fig:tc_more}%
            (Color online) The critical temperature as function of the thickness in the case of different metal overlayers.}
\end{figure}

The theory presented in this paper so far  can be applied without changes to other overlayer systems like NbAg, NbIr, NbAl and NbMo.
Nevertheless, it should be mentioned, that many of the properties of the quasiparticle spectrum we studied above 
are connected to the fact that QW states form in the Au in the normal state. The formation of QW states however, can not be regarded as a universal feature 
of every overlayer on Nb. This is illustrated in 
Fig.~\ref{fig:not_qw}, where we show the normal state band structure and also the corresponding quasiparticle spectrum of
Al and Mo overlayers on the Nb host. While there may be some quantum confinement present in these materials in the normal state, 
it is not as clearly exhibited as in the case 
of Au overlayers. It is clearly seen, that quantum confinement is still present in the superconducting state. 
It is important to notice as well, that in these systems it is hard to recognize how the quasiparticle specrtum could be obtained from the normal state band structure, as we
mentioned before, based on the symmetry properties of the BdG equations.
The most important feature of these plots is that in the superconducting state the induced gap around the Fermi-level can still be observed, therefore, 
the  superconducting transition temperature can be predicted similarly to the NbAu heterostructure. This is shown in Fig.~\ref{fig:tc_more} for several overlayer systems.

It can be seen that it is a general trend that the superconducting transition temperature decays as the normal metal thickness 
increases, only the rate of the decay depends on the material of the overlayer.
This behavior is a consequence of the size of the induced gap in the normal
metal and also the change of the density of states at the Fermi level which enters the McMillan formula
through the effective electron-phonon coupling parameter.
This is most pronounced for the case of iridium overlayers, where the decay of the transition temperature is the fastest, mostly due to the
much higher density of states at the Fermi level.

In summary, we have investigated the quasiparticle spectrum, and the superconducting
transition temperature of superconductor -- normal metal heterostructures.
We have developed a method to predict the transition temperature of such heterostructures based on the first-principles solution of the
KS-BdG equations.  In the case of the Nb/Au(100) system we obtained very good agreement with the experimental findings.
The theory was also applied for different metallic overlayers on a Nb host to predict the superconducting transition
temperature.

{\emph{Acknowledgment}} ---  This work was supported by the Hungarian  National, Research, Development and Innovation Office
under the contracts No. K115632 and K108676.

\bibliographystyle{apsrev4-1}
\bibliography{tcpaper}

\end{document}